\documentclass[twocolumn,10pt,aps,prl,nofootinbib]{revtex4-1}

\usepackage{amssymb, amsmath, amsthm}
\usepackage{graphicx}
\usepackage{units}

\begin{document}

\title{Optical response of gas-phase atoms at less than $\lambda/80$ from a dielectric surface}

\author{K. A. Whittaker$^{1}$}
\author{J. Keaveney$^{1}$}
\author{I. G. Hughes$^{1}$}
\author{A. Sargsyan$^{2}$}
\author{D. Sarkisyan$^{2}$}
\author{C. S. Adams$^{1}$}
\affiliation{$^{1}$ Joint Quantum Centre (JQC) Durham-Newcastle, Department of Physics, Durham University, South Road, Durham, DH1 3LE, United Kingdom}
\affiliation{$^{2}$ Institute for Physical Research, National Academy of Sciences - Ashtarak 2, 0203, Armenia}

\date{\today}

\begin{abstract}
We present experimental observations of atom-light interactions
within tens of nanometers (down to 11~nm) of a sapphire surface. Using photon counting we detect the fluorescence from of order one thousand Rb or Cs atoms, confined in a vapor with thickness much less than the optical excitation wavelength. The asymmetry in the spectral lineshape provides a direct read-out of the atom-surface potential. A numerical fit indicates a power-law $-C_{\alpha}/r^{\alpha}$ with $\alpha=3.02\pm0.06$ confirming that the van der Waals interaction dominates over other effects. The extreme sensitivity of our photon-counting technique may allow the search for atom-surface bound states.
\end{abstract}

\maketitle

Atomic vapors are continuing to find new applications in quantum technologies such as chip-scale atomic clocks~\cite{Knappe2004};
magnetometry~\cite{Griffith2009,Mhaskar2012}; 
magnetoencephalography~\cite{Sander2012};
an atom-based optical isolator~\cite{Weller2012b};
quantum memories~\cite{Julsgaard2004};
frequency filtering~\cite{Abel2009,Zielinska2012};
and in the field of nanoplasmonics (see refs.~\cite{Gramotnev2010,Stockman2011} for reviews). As the miniaturization of these technologies progresses, many of these systems eventually reach the scale where the proximity of the atoms to a surface becomes significant.
In this case a thorough understanding of the atom-surface interactions is essential.
Many of the above applications use atoms in ground states or low-lying excited states, where the atom-surface (AS) interaction is relatively small as the induced dipole is only a few Debye. 
Even so, the AS interaction can still have a significant effect if the surface is in the near-field of the atom, that is, within a fraction of the transition wavelength, $\lambda$, of the induced dipole.
In this regime the atom-surface potential is governed by an inverse power law $U_{\rm vdW} = -C_{\alpha}/r^{\alpha}$ where ${C_{\alpha}}$ is the coupling coefficient and ${r}$ is the atom-surface distance. 
For an uncharged surface with $r<\lambda$ one expects a van der Waals interaction with $\alpha=3$~\cite{Casimir1948}.
However, if charges are present on the surface the Coulomb interaction may be larger than the van der Waals interaction, leading to a modification of $\alpha$.
The atom-surface potential is also strongly influenced by the presence of surface modes such as surface polaritons. 
However, for alkali atoms these couple more strongly to intermediate excited states where the energy level spacing is in the Terahertz region~\cite{Failache1999,Failache2002,Failache2003,Kubler2010,Gonzalez-Tudela2013}. 
Very close to the surface, bound states of the AS potential can be exploited, as recently demonstrated using He scattering from LiF surfaces~\cite{Debiossac2014}.
\begin{figure}[b]
\centering
 \includegraphics[width=0.49\textwidth,angle=0]{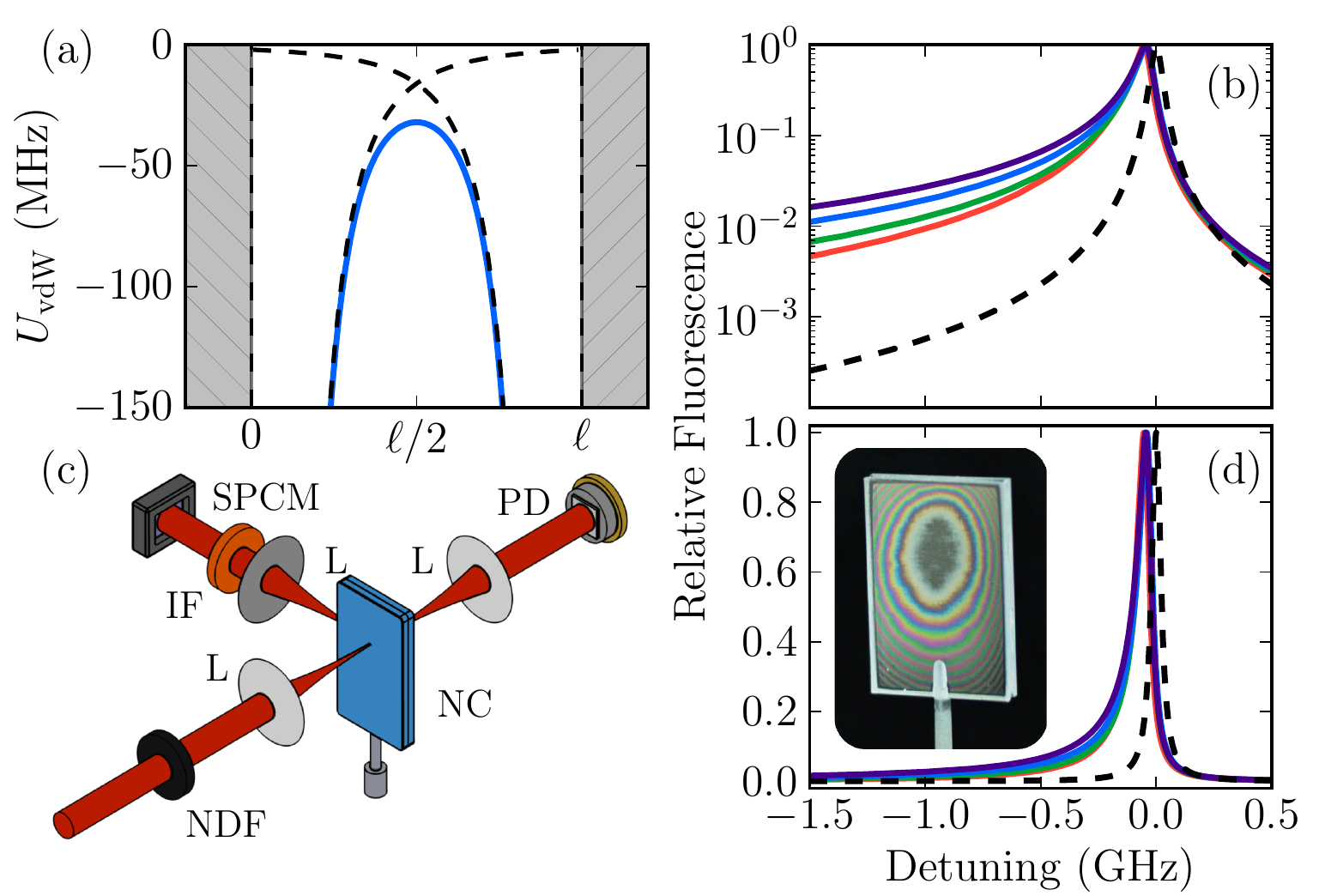}
 \caption{(a) Atom-surface potential in the nanocell (NC) due to both walls (dashed lines) and combined (blue line). (b/d) The effect of the atom-surface interaction on an initially Lorentzian spectral line (black dashed curves) is to shift the peak position, and create a pronounced asymmetry between the red and blue wings (red curves). The amount of asymmetry is a direct readout of the exponent $\alpha$, as demonstrated by the four solid lines which are calculated with $\alpha=4,3,2,1.5$ (top to bottom). Panel (c) shows a schematic of the experimental setup used to detect off-axis fluorescence. NDF - neutral density filter; L - lens; NC - nanocell; IF - interference (bandpass) filter; PD - photodiode; SPCM - single photon counting module. Finally, a photo of the Cs nano-cell used in the experiment is shown in the inset to panel (d). At the center of the Newton's rings interference pattern the thickness of the vapor column is 50~nm.}
 \label{fig:setup}
\end{figure}
The combination of bound states and surface resonances potentially allows guiding or trapping of atoms in close proximity to the surface~\cite{Chang2013a}. 
This could lead to a new type of hybrid nanoscale atom-surface metamaterials, with atoms trapped in small channels that can be etched into any conceivable geometry, using focussed ion beam milling, for example~\cite{Baluktsian2010}.

The atom-surface interaction may be studied using a variety of methods. Scattering or deflection of an atom beam from a metallic surface~\cite{Sandoghdar1992,Sukenik1993,Shimizu2001,Druzhinina2003,Pasquini2004,Bender2010}; 
deflection of an ultra-cold atomic cloud from an atomic mirror~\cite{Landragin1996,Mohapatra2006c} or
diffraction of an atomic beam~\cite{Grisenti1999,Perreault2005} have all been demonstrated. In these examples detection occurs after the interaction has taken place. 
For measurements of near-field effects at specific length scales such as atomic guiding, real-time in-situ detection is preferable. Spectroscopic studies can be used, though at the cost of probing the difference in AS interaction between two atomic states. Considerable insight has been gained using of frequency-modulated selective reflection (FMSR) spectroscopy in atomic vapors~\cite{Oria1991,Hamdi2005,Fichet2007a,Laliotis2007a}, which probes the vapor with a distance of order $\lambda$ from the surface.
Although FMSR is useful in determining the average shift from zero-crossings, 
extracting detailed information from the lineshape is complicated by the effects of dipole-dipole interactions between atoms, leading to self-broadening \cite{Weller2011a} and shifts \cite{Keaveney2012}, and for parallel surfaces the windows act as a low-finesse etalon which adds further complication to transmission and selective reflection signals~\cite{Dutier2003a,Keaveney2013}.

In this work we detect fluorescence from an atomic vapor with nanoscale thickness, and use photon counting to probe the atom-surface interaction at low atomic density where other interactions are negligible. 
This yields optimum resolution of the spectral lineshape, which is a direct probe of the AS potential. 

By fitting this high-resolution data to a comprehensive model of the atomic susceptibility, which has previously been used to model transmission and refraction in thermal temperature vapor cells in a range of experimental regimes~\cite{Siddons2008b,Weller2011a,Keaveney2012,Keaveney2012a,Weller2012,Weller2012a,Weller2012b,Zentile2014},
we extract the atom-surface interaction and thereby calibrate the position of an atom emitting at a particular frequency. Using this method, we are able to detect atoms within 10-15 nm of the dielectric surface. 
The possibility to exploit this length scale opens interesting prospects for strong coupling between atoms and nanoscale plasmonic structures or localized polaritons~\cite{Chang2013a}.  

Figure~1 illustrates the general principle of our experiment, and the expected fluorescence lineshapes. By confining the vapor between two surfaces, the interaction volume and hence the distribution of van der Waals shifts, as shown in Fig.~1(a), is independent of other variables such as the atomic density. 
The overall potential $U_{\rm vdW}$ (red solid line) is the sum of both surfaces (dashed lines), and so has a minimum at the center of the cell, which quickly diverges at either surface.
In the absence of the AS interaction, the fluorescence lineshape is best approximated as a symmetric Lorentzian function (owing to Dicke narrowing, the fluorescence lineshape is not the usual Gaussian seen in conventional thermal vapor cells). The AS interaction causes an asymmetry in the fluorescence lineshape, as shown in Fig.~1(b/d), and therefore gives a direct read-out of the AS potential.
This is most striking when shown on a log scale (panel (b)), though is also evident on a linear scale, where the shift of the peak is more obvious. 
\begin{figure}[t]
\centering
 \includegraphics[width=0.48\textwidth,angle=0]{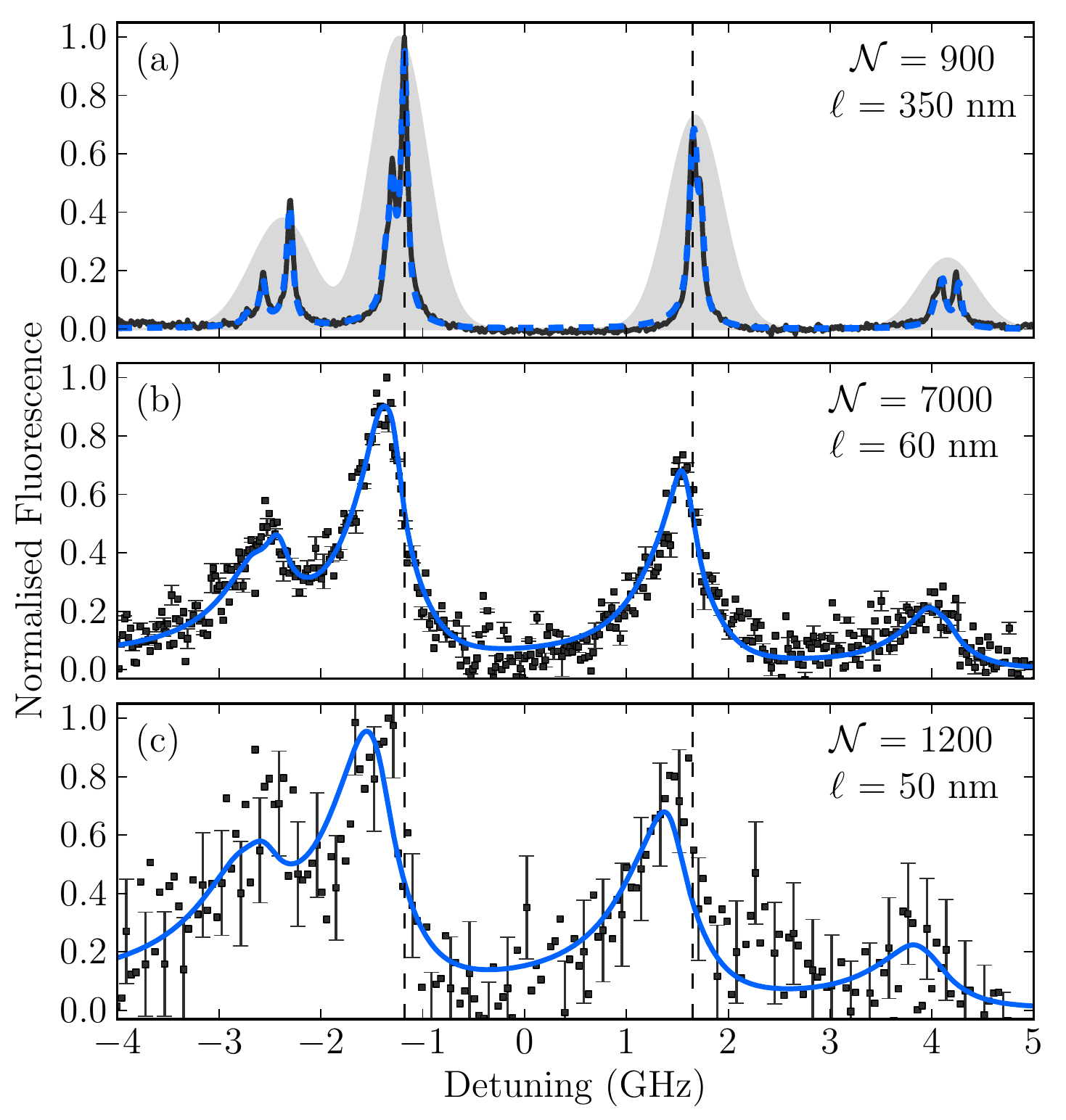}
 \caption{(a) Fluorescence from a Rb vapor with thickness $\ell = 350$~nm at a temperature $T=85^{\circ}$C, integration time approximately 15~minutes. At this thickness the atom-surface interaction is negligible and owing to strong Dicke narrowing, the lineshape of each hyperfine transition is well approximated by a Lorentzian with a (fitted) FWHM of $59 \pm 1$~MHz, and it is therefore possible to resolve each hyperfine transition. For comparison, the the grey outline shows the fluorescence signal expected from a conventional Doppler-broadened vapor. 
 (b) Fluorescence from a vapor with thickness $\ell = 60$~nm at a temperature $T=150^{\circ}$C, integration time approximately 1~hour. The atom-surface interaction significantly shifts the center of the fluorescence peaks, and because of the decreased time-of-flight of the atoms the lines are broadened. However, the linewidth is still narrower than the Doppler width.
 (c) Fluorescence from a vapor with thickness $\ell = 50$~nm at a temperature $T=130^{\circ}$C, integration time approximately 8~hours. Due to the thickness of the vapor and the reduced number density, the signal-to-noise (SNR) is much weaker than in panel (b), even though the data has been binned more coarsely. However, a good fit to our theoretical model is still achieved. In this panel the asymmetry between red- and blue-detuned wings is most pronounced.
In all panels, $\cal{N}$ is the mean number of atoms probed at any one time by the laser. Zero detuning represents the weighted center of the Rb D2 absorption line.}
 \label{fig:Rb}
\end{figure}
The experimental set-up and the cell are illustrated in Fig.~1(c) and 1(d) inset, respectively. The laser is scanned across resonance and fluorescence photons are counted on a single photon counting module to acquire a spectrum. In addition to the off-axis fluorescence, we also detect transmitted light, however the amount of absorption at the densities considered in this work is small, typically $<0.1$\%, so no useful signal is obtained here.
The probe laser power is around 1~$\mu$W and the beam is focussed to a waist ($1/e^{2}$ radius) of 30~$\mu$m inside the cell. The background counts due to thermal photons from the heater accumulate at a constant rate and are subtracted from the data during post-processing.

For spectroscopic reference and calibration, we also monitor the transmission of the laser light through a 7.5~cm vapor cell, and linearise the laser scan using a Fabry-Perot etalon, in the same way as our previous work~\cite{Siddons2008b,Weller2011a,Keaveney2012}.
More details on the photon counting technique can be found in ref.~\cite{Keaveney2013}.

\begin{figure}[t]
\centering
 \includegraphics[width=0.48\textwidth,angle=0]{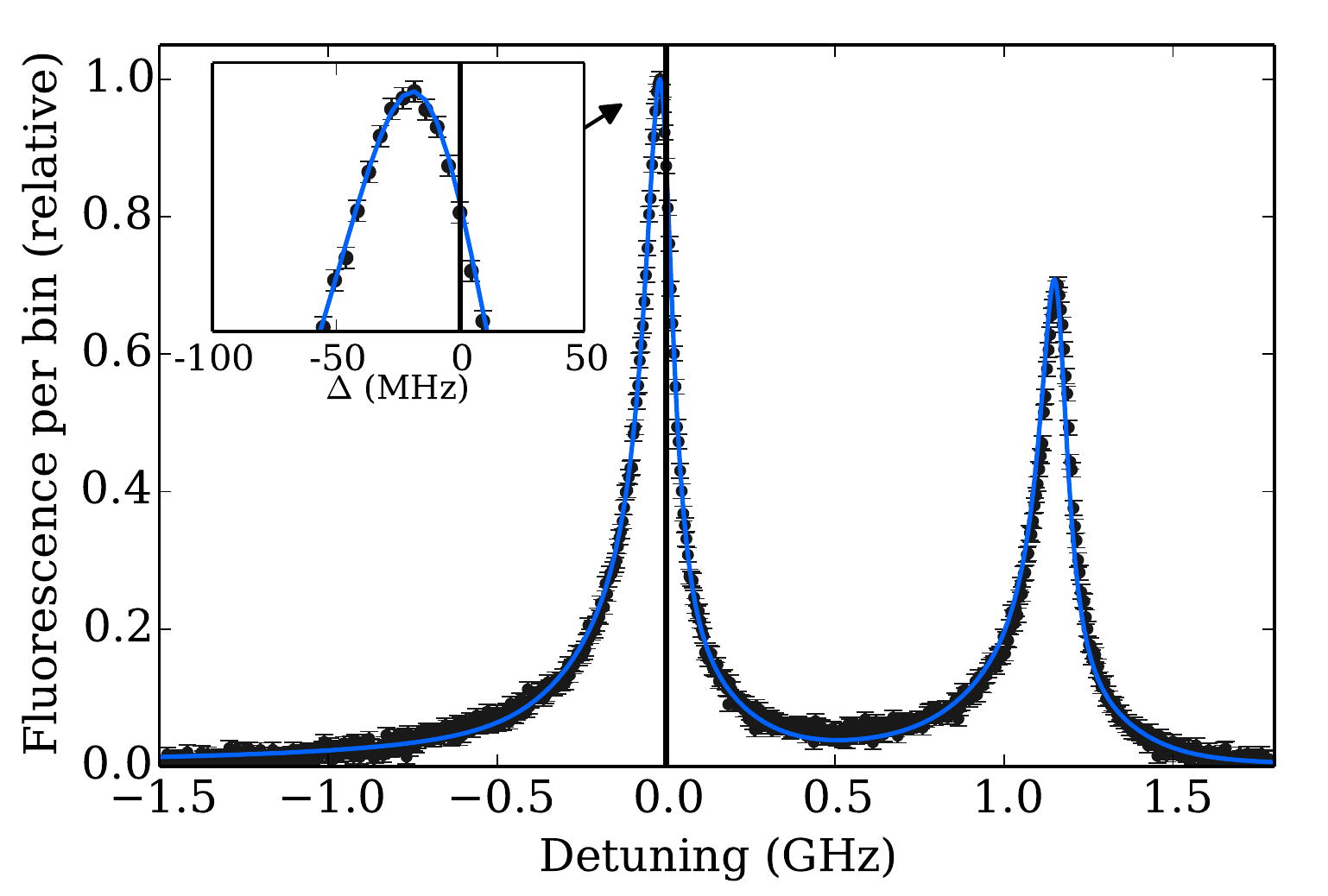}
 \caption{Fluorescence from a Cs vapor with thickness $\ell = 150$~nm at a temperature $T=140^{\circ}$C, integration time approximately 8~hours. At this thickness the atom-surface interaction is small ($20\pm1$~MHz peak shift), but measurable with this detection technique. Fitting these data to our model yields an excellent fit and from this we are able to extract a C$_3$ coefficient for the AS interaction (see main text). The Dicke narrowing is still particularly striking at this thickness - even though the atoms are at most 75~nm from the surface, the fitted linewidth is only $86\pm1$~MHz.
Zero detuning is the resonance frequency of the Cs 6S$_{1/2} \ F_{\rm g}=4\rightarrow$ 6P$_{1/2} \ F_{\rm e}=3$ transition. }
 \label{fig:Cs}
\end{figure}

Figure~\ref{fig:Rb} shows fluorescence spectra in Rb vapor at various cell thicknesses. In panel (a) we show fluorescence from a vapor with thickness $\ell = 350$~nm, at a temperature $T=85^{\circ}$C which corresponds to an atomic number density $N = 2 \times 10^{12}$~cm$^{-3}$. 
With this density and our beam geometry the laser interacts with on average around $\mathcal{N}=900$ atoms at any one time; the vapor has a peak optical density~$\sim 5 \times 10^{-3}$. 
Compared with the normal Doppler profile (grey area), the spectra are considerably narrower. After fitting to our model we extract a Lorentzian linewidth of $59 \pm 1$~MHz.
At $\ell = 350$~nm (b) the AS interaction is negligibly small. In panels (b) and (c) we present data where the AS interaction is significant. At $\ell=60$~nm spectral broadening due to reduced time-of-flight and dipole-dipole interactions (self-broadening) impairs resolution of individual hyperfine resonances, but a shift of the spectral features is noticeable, and is most pronounced on the two strongest ($^{85}$Rb) spectral features.
At $\ell = 50$~nm (c), we reach the minimum width at which it is still possible to obtain a reasonable fit to the model. Here we use a lower temperature to reduce dipole-dipole interactions, but this comes at the cost of reduced SNR. 
However, we still achieve a good fit. As the atoms are never more than 25~nm from a surface, both the shift and asymmetry of the spectral lines due to the AS interaction are significant.

Whilst in principle one could extract a C$_{3}$ coefficient from the shift of the peak position relative to the unshifted position (black dashed line) this is difficult due to the many overlapping lines. A better approach is to fit the full spectral profile, which we do by fitting to our model with a floating C$_{3}$ parameter. To compute the lineshape, we convolve the surface potential with the atomic response in the absence of a surface potential, and assuming that the atomic interactions are uniformly distributed over the cell. 

We vary the cell thickness from 50~nm to 390~nm ($\lambda/2$) to obtain many data sets like the ones in figure~\ref{fig:Rb}. By fitting all of the data with shared parameters, we extract an AS interaction strength C$_{3} = 1.2 \pm 0.3$~kHz~$\mu$m$^{3}$ for the Rb 5S$_{1/2}$~$\rightarrow$~5P$_{3/2}$ transition. 
This is in reasonable agreement with a theoretical value of 1.8~kHz~$\mu$m$^{3}$ (taken from ref.~\cite{Derevianko1999} correcting for the surface reflectivity), and the relatively large error bar is probably due to the fact that each hyperfine transition has a slightly different dipole moment and thus a slightly different C$_{3}$ coefficient. However, fitting the data with 12 free interaction parameters instead of just one is computationally infeasible.
Fitting these data with a surface potential of the form $-C_{\alpha}/r^{\alpha}$, where $\alpha$ is a floating parameter, allows the verification of the expected van der Waals $r^{-3}$ power-law. 
Fitting all our data we extract a weighted average $\alpha = 3.02 \pm 0.06$, confirming that there are no surface charges or other contaminants, and that the AS interaction follows the expected van der Waals form.

In contrast to the complexity of the Rb D2 line, the Cs D1 line is much simpler. In this system, the increased hyperfine splitting and the presence of only one isotope makes possible the investigation of individual hyperfine transitions. The ground state hyperfine splitting is over 9~GHz and the excited state hyperfine splitting is 1.17~GHz, still more than the Doppler width. Example data are shown in figure~\ref{fig:Cs}, for a cell thickness $\ell = 150$~nm at a temperature $T=140 \; ^{\circ}$C, corresponding to a Cs atomic density $N=1.4\times 10^{14}$~cm$^{-3}$.
At this density, we expect the dipole-dipole interactions between atoms to contribute 11~MHz to the total linewidth. We extract from the fit a total Lorentzian linewidth of just $(86\pm1)$~MHz, and attribute the additional width to a time-of-flight broadening due to the cell geometry.

The relative narrowness of the peaks and good SNR means that we can detect shifts on the order of a few MHz with high precision, as shown in the inset of figure~\ref{fig:Cs}. In this case the shift of the peak is $(20\pm 1)$~MHz, where the error bar is based on fitting just the peak of the data to a Lorentzian.
As we scan over only two transitions, the full spectral analysis is much simpler, and for the data shown we achieve a reduced $\chi^{2}$ parameter of 1.75, indicating an excellent fit~\cite{Hughes2010}.

\begin{figure}[t]
\centering
 \includegraphics[width=0.46\textwidth,angle=0]{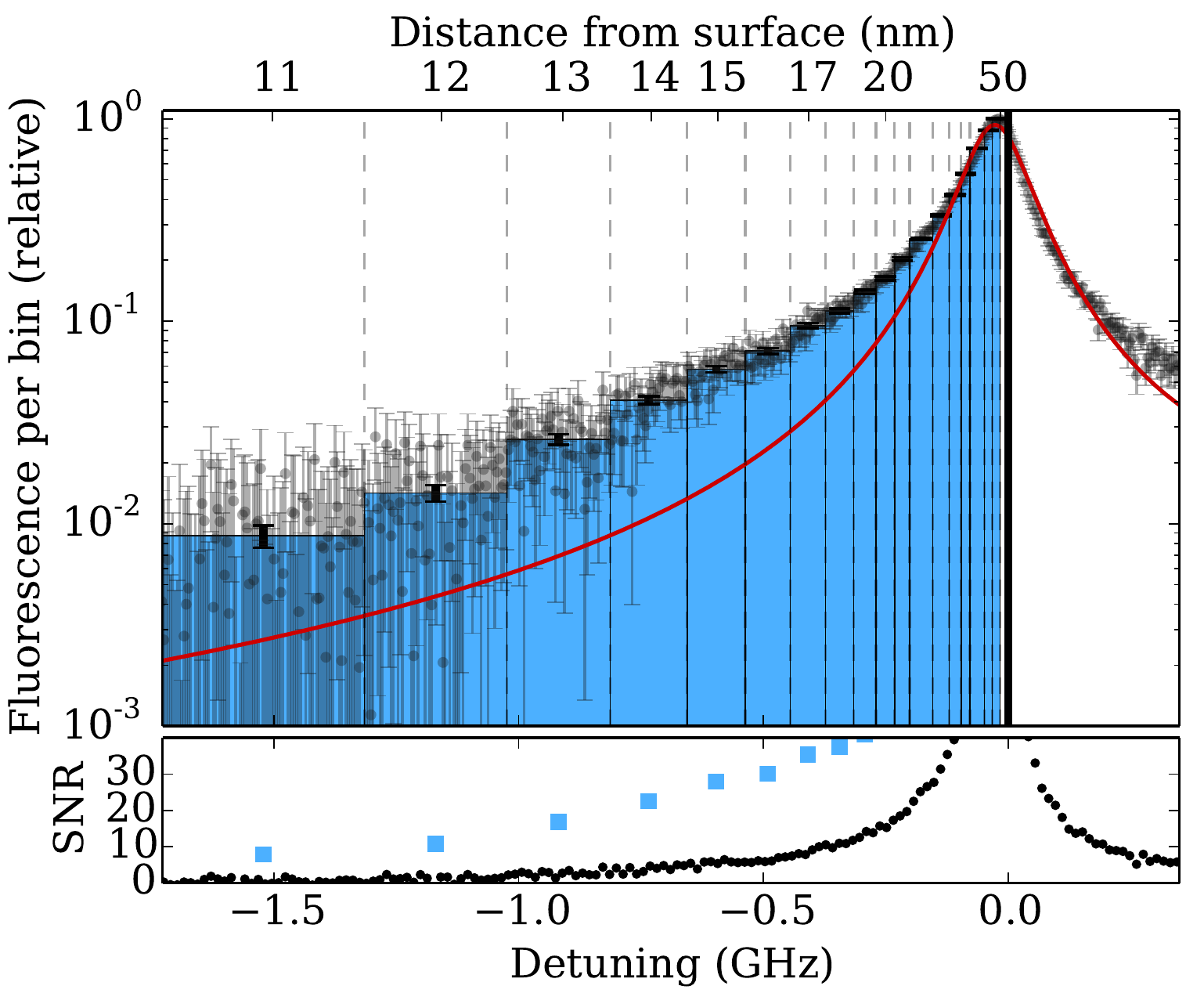}
 \caption{The grey points and error bars show the same fluorescence data as figure~\ref{fig:Cs}, but shown on a logarithmic scale. Whereas the asymmetry in the lineshape may not be apparent from figure 3, when plotted on a log scale the difference from a Lorentzian (red curve) that is fitted to the blue-wing of the data is striking. 
The fluorescence at any given laser frequency detuning maps directly onto the distance of an atom from (either) one of the surfaces of the cell.  
By processing the photon arrival data into (uneven) frequency bins (blue bars, normalised by population density) whose width represents a distance $\pm0.5$nm from the surface, we increase the effective signal-to-noise ratio (SNR) over the normal evenly binned data (grey points). These data are clearly above that which would be expected from a standard Lorentzian resonance line (solid red line), and confirms the detection of atoms within 11~nm of a sapphire surface. For these data we are limited by the scan range of the laser, not SNR.}
 \label{fig:SNR}
\end{figure}

Though the shift is clear from figure~\ref{fig:Cs}, the asymmetry in the lineshape is not particularly apparent. In figure~\ref{fig:SNR} we plot the same data on a logarithmic scale. The deviation in the red wing from the symmetric Lorentzian (red line) is immediately apparent. The small deviation on the blue wing is due to the other hyperfine transition at a detuning of 1.17~GHz.

Since the AS interaction maps atomic position to a frequency shift, we can interpret the fluorescence data as a direct readout of the atomic position in the cell. On the alternate axis in figure~\ref{fig:SNR} we bin the data into non-uniform frequency steps such that the width of one bin corresponds to an atomic position change of 1~nm. Because the bins are so large in the wing of the resonance, this dramatically increases the SNR. From this, we can conclusively detect atoms $(11.0\pm0.5)$~nm away from a surface, the signal level is many standard errors above that expected from the normal Lorentzian wing. 
Using the same procedure employed for Rb, we take a range of data where we vary the cell thickness between 80~nm and 200~nm and fit all the data with a combined C$_{3}$ parameter. 
From this, we extract a spectroscopic $C_{3}$ of $(1.9\pm0.1)$~kHz~$\mu$m$^{3}$ between sapphire and the Cs 5S$_{1/2}\rightarrow$~6P$_{1/2}$ transition, in agreement with previous work~\cite{Oria1991}.

The reduced detection efficiency owing to the wavelength-dependent quantum efficiency of the SPCMs and quality of bandpass filters available means that detecting Cs fluorescence is technically more difficult. The increased sensitivity to thermal photons produced in the cell heater also limits the maximum atomic density that is feasible to investigate using the current equipment. However, the main limit in the current experiment is how far we can scan the laser, not SNR. In future work we will investigate the region farther out in the red-detuned wing and look for a signature of atom-surface bound-states, which have been predicted to occur at a detuning of around -20~GHz~\cite{Lima2000}.

In conclusion, we have demonstrated a simple method for in-situ detection of atoms a small fraction of a wavelength away from a dielectric surface, and used this method to investigate the AS interaction between sapphire and both Rb and Cs vapors in their first excited states. The spectral lineshape is directly connected with the surface potential, and by analysing the spectroscopic data we have both confirmed the expected $1/r^{3}$ power-law and from this calculated the C$_{3}$ interaction strength coefficients for Rb and Cs. This technique could be used to probe long-range atom-surface bound states~\cite{Lima2000}, or for the detection of atoms confined in nano-cavities which could find application as part of a micromechanical resonantor system, similar to those in ref.~\cite{Safavi-Naeini2013}. These topics will form the basis of future research.

The authors would like to thank K. N. Jarvis for experimental assistance, and acknowledge financial support from EPSRC, grant reference EP/H002839/1. The data presented in this paper are available upon request. Correspondence should be addressed to either james.keaveney@durham.ac.uk or c.s.adams@durham.ac.uk.

\bibliographystyle{apsrev4-1}
\bibliography{library}

\end{document}